# Unidirectional spectral singularity lasing in a defective atomic lattice

Chen Peng,[1] Xinfu Zheng,[1] Duanfu Chen,[1] Hanxiao Zhang,[1] Dong Yan,[1] Jinhui Wu,[2,*] and Hong Yang[1,†]

[1] *School of Physics and Electronic Engineering, Hainan Normal University, Haikou 571158, People's Republic of China*

[2] *School of Physics and Center for Quantum Sciences, Northeast Normal University, Changchun 130024, People's Republic of China*

We propose an efficient scheme for achieving mode-tunable unidirectional reflection lasing (URL) by establishing a coherent gain atomic system to amplify the probe field and ingeniously designing the one-dimensional (1D) defective atomic lattice. This lattice not only replaces the resonant cavity to provide a distributed feedback mechanism but also breaks the spatial symmetry of the probe susceptibility. Correspondingly, the URL can be characterized by a non-Hermitian degenerate spectral singularity (NHDSS), where the two eigenvalues of the inverse scattering matrix are engineered to satisfy $\lambda^{+}_{S^{-1}} \simeq \lambda^{-}_{S^{-1}} \to 0$. This intriguing NHDSS depends on the probe susceptibility and the Bragg condition, both of which can be modulated by adjusting the external optical field and lattice structure, rendering the scheme experimentally feasible. Our approach achieves both nonreciprocity and lasing oscillation in a single system, significantly enhancing the efficiency of optical information transmission and facilitating the integration of active photonic devices into compact quantum networks.

## I. INTRODUCTION

The nonreciprocal laser has significant applications in quantum networks and optical information processing due to its high efficiency, high density and narrow linewidth quantum entanglement [1–3], which is usually built on the basis of nonreciprocal amplification. Nonreciprocal amplification [4–6] enables the amplification of sensitive and fragile weak signals in quantum systems, while isolating them from external noise induced by backscattering, [7–10], a phenomenon that has been extensively investigated in numerous active systems [11–15]. However, the realization of nonreciprocal lasers not only requires a gain medium but also relies on an optical resonator. In recent years, extensive research has been conducted on realizing nonreciprocal lasers with the aid of optical resonators. For instance, in a system composed of a spinning microwave resonator coupled to a magnetic material, nonreciprocal phonon laser can be realized by leveraging the Fizeau light-dragging effect [16], alternatively, the optical Sagnac effect can be exploited in a coupled cavity system, which consists of an optomechanical resonator and a spinning resonator, to achieve the same result [17]. Besides, directional phonon lasers can be realized by coupling optomechanical resonators to nonlinear optical resonators [18]. This type of nonreciprocal phonon device, characterized by directional amplified phonon flow, can be applied to chiral phonon transport, phonon isolation, unidirectional mechanical networks, and invisible acoustic sensing, among others [19–22]. Similarly to the rapid development of phonon lasers, magnon lasers utilizing magnon-induced Brillouin scattering have seen rapid progress. The manipulation of nonreciprocal magnon lasers can be effectively achieved through the Fizeau light migration effect in both optical cavity [23–26] and parity-time (PT)-symmetric cavity optomagnonic systems [27, 28], where the effect is caused by resonator rotation.

It can be seen that the resonator cavity plays a crucial role in both phonon lasers and magnon lasers. Similarly, the realization of nonreciprocity photonic lasers is generally based on resonator cavity systems, where optical signals achieve the threshold condition through repeated oscillation to output lasing, followed by the implementation of nonreciprocity via symmetry-breaking effects. Currently, research on nonreciprocal photonic lasers primarily focuses on systems such as micro-ring resonators and silicon waveguides [29, 30], coupled-cavity atomic ensembles [31] , Josephson ring systems [32], Reservoir engineering [33], chiral materials [34] and non-Hermitian time-Floquet systems [35]. To develop high-precision lasers, the cold-atom laser [36–39] has been favored by scholars due to their coherence and low relative noise intensity. Nonreciprocal lasing has been observed in a ring cavity using cold atomic gas as the gain medium, which produces a narrow-band lasing in $^{88}$Sr atom using collective effect [40]. However, the periodic structure of an optical lattice can be used to realize distributed feedback as a substitute for the resonator cavity [41–43], which, when combined with a gain atomic system, can achieve two-color reciprocal lasing output [38]. In addition, based on FWM gain, the optical parametric amplification based lasing emission has been achieved experimentally in the optical lattice trapping cold $^{87}$Rb atoms [44]. In the work [45, 46], we have realized the nonreciprocal reflection by breaking the spatial symmetry of probe susceptibility through lattice defects. And the experimental results of light propagation in the atomic lattice show ex-cellent agreement with theoretical predictions [47, 48]. Furthermore, we have achieved unidirectional reflection amplification by combining an active atomic system with a defective atomic lattice [49]. However, how to achieve nonreciprocal lasing in defective atomic lattice systems?

* jhwu@nenu.edu.cn
† yang_hongbj@126.com

In this letter, we propose an efficient scheme for achieving mode-tunable unidirectional reflection lasing (URL) in a one-dimensional (1D) defective atomic lattice assisted by the gain atomic system, by leveraging non-Hermitian degenerate spectral singularity (NHDSS) that the two eigenvalues of its scattering matrix engineered to be $\lambda^{+}_{S^{-1}} \simeq \lambda^{-}_{S^{-1}}$ (indicates the unidirectional reflection) [50–56] and $\lambda^{+}_{S^{-1}} \to 0$ (means generating lasing output) [57–61]. The physical essence lies in the fact that the 1D defective atomic lattice can break the spatial symmetry of the probe susceptibility and provide a distributed feedback mechanism. Subsequently, the probe light amplified by the gain atomic system undergoes multiple oscillations via sufficient Bragg scattering induced by the distributed feedback, ultimately being enhanced into a lasing output. In particular, we can dynamically control the position, intensity and line width of URL through the modulation of relevant parameters, and provide an experimental setup.

## II. THEORETICAL MODEL AND EQUATIONS

As shown in Fig. 1 (c), 1D defective atomic lattice is divided into two parts, part I with $p_1$ periods, each period consisting of $a_f$ filled cells (trapped atoms) and $a_v$ vacant cells (without atoms), and part II with $p_2$ periods of filled cells. The total length of the lattice is $L = S\lambda_0$, where $S = (a_f + a_v)p_1 + p_2$. The atoms are prepared in a three-level configuration by a weak probe field $\mathbf{E_p}$, a strong coupling field $\mathbf{E_c}$ and a microwave field $\mathbf{E_w}$, as shown in Fig. 1(a). The $D_1$ line of $^{87}$Rb atoms is chosen here with $|1\rangle = |5S_{1/2}, F = 2, m_F = -2\rangle$, $|2\rangle = |5S_{1/2}, F = 2, m_F = 0\rangle$ and $|3\rangle = |5P_{1/2}, F = 2, m_F = -1\rangle$. We define the Rabi frequency of the probe field $\Omega_p = \mathbf{E_p} \cdot \mathbf{d_{13}}/2\hbar$ with the detuning $\Delta_p = \omega_{31} - \omega_p$, the Rabi frequency of the coupling field $\Omega_c = \mathbf{E_c} \cdot \mathbf{d_{23}}/2\hbar$ with the detuning $\Delta_c = \omega_{32} - \omega_c$ and the Rabi frequency of the resonant microwave field $\Omega_w = \mathbf{E_w} \cdot \mathbf{d_{12}}/2\hbar$ with the detuning $\Delta_w = \omega_{21} - \omega_w = 0$. The matrix element $\mathbf{d}_{ij} = \langle i|\,\mathbf{d}\,|j\rangle$ is used to denote the dipole moment of transition $|i\rangle$ to $|j\rangle$.

Within the electric-dipole and rotating-wave approximations, the total Hamiltonian of the system which describes the atom-field coupling can be expressed as follows:

$$\begin{aligned}\hat{H}_{tot} =& \omega_{21}\hat{\sigma}_{22} + \omega_{31}\hat{\sigma}_{33} - (\Omega_{w_0} e^{-i\omega_w t}\hat{\sigma}_{21} \\ &+ \Omega_{p_0} e^{-i\omega_p t}\hat{\sigma}_{31} + \Omega_{c_0} e^{-i\omega_c t}\hat{\sigma}_{32} + \text{H.c.}),\end{aligned} \tag{1}$$

where the corresponding Rabi frequencies are redefined as $\Omega_{p_0} = \Omega_p e^{i\phi_p}$, $\Omega_{c_0} = \Omega_c e^{i\phi_c}$ and $\Omega_{w_0} = \Omega_w e^{i\phi_w}$. Here $\Omega_p$, $\Omega_c$ and $\Omega_w$ are real numbers. Based on the total Hamiltonian Eq. (1), by choosing the external classical optical-field frequency, we can perform a unitary transformation to a rotating coordinate frame described by the unitary operator $\hat{U} = e^{-i\hat{H}_f t}$, where $\hat{H}_f = (\omega_p - \omega_c)\,\hat{\sigma}_{22} + \omega_p\hat{\sigma}_{33}$. In terms of the formula $\hat{H}_{rot} = \hat{U}^{\dagger}\hat{H}_{tot}\hat{U} - i\hat{U}^{\dagger}(\partial\hat{U}/\partial t)$, we obtain time-independent Hamiltonian, namely

$$\begin{aligned}\hat{H}_{rot} =& (\Delta_p - \Delta_c)\,\hat{\sigma}_{22} + \Delta_p\hat{\sigma}_{33} - (\Omega_{w_0}\hat{\sigma}_{21} + \Omega_{p_0}\hat{\sigma}_{31} \\ &+ \Omega_{c_0}\hat{\sigma}_{32} + \text{H.c.}),\end{aligned} \tag{2}$$

without loss of generality, we adopt the same method as in [62], thus the atomic transformations are redefined as $\hat{\sigma}_{31} \to \hat{\sigma}_{31} e^{-i\phi_p}$, $\hat{\sigma}_{32} \to \hat{\sigma}_{32} e^{-i\phi_c}$ and $\hat{\sigma}_{21} \to \hat{\sigma}_{21} e^{i(\phi_c - \phi_p)}$. In the closed-loop configuration, the phases of the probe field, coupling field and microwave field are linked, making the relative phase $\phi = \phi_w + \phi_c - \phi_p$ a key factor. Based on this transformation and considering the incoherent (dissipative) processes, the density-matrix operator $\hat{\rho}$ of the atomic system is described by the Lindblad master equation in the Born-Markov approximation:

$$\begin{aligned}\partial_t\hat{\rho} =& -i\left[\hat{H}_{tra}, \hat{\rho}\right] + \Gamma_{21} e^{i\phi}\mathcal{L}(\hat{\sigma}_{12})\,\hat{\rho} \\ &+ \Gamma_{31}\mathcal{L}(\hat{\sigma}_{13})\,\hat{\rho} + \Gamma_{32}\mathcal{L}(\hat{\sigma}_{23})\,\hat{\rho},\end{aligned} \tag{3}$$

with

$$\begin{aligned}\hat{H}_{tra} =& (\Delta_p - \Delta_c)\,\hat{\sigma}_{22} + \Delta_p\hat{\sigma}_{33} - (\Omega_w e^{i\phi}\hat{\sigma}_{21} \\ &+ \Omega_p\hat{\sigma}_{31} + \Omega_c\hat{\sigma}_{32} + \text{H.c.}),\end{aligned} \tag{4}$$

the above Lindblad superoperator $\mathcal{L}(\hat{\sigma})$ describes the dissipative coupling to the environment and is given by the form $\mathcal{L}(\hat{\sigma})\,\hat{\rho} = \hat{\sigma}\hat{\rho}\hat{\sigma}^{\dagger} - \hat{\sigma}^{\dagger}\hat{\sigma}\hat{\rho}/2 - \hat{\rho}\hat{\sigma}^{\dagger}\hat{\sigma}/2$. Finally, the density-matrix equations are written as follows:

$$\begin{aligned}\partial_t\hat{\rho}_{11} =& i\Omega_p(\hat{\rho}_{31} - \hat{\rho}_{13}) + i\Omega_w(\hat{\rho}_{21}e^{-i\phi} - \hat{\rho}_{12}e^{i\phi}) \\ &+ \Gamma_{31}\hat{\rho}_{33}, \\ \partial_t\hat{\rho}_{22} =& i\Omega_c(\hat{\rho}_{32} - \hat{\rho}_{23}) - i\Omega_w(\hat{\rho}_{21}e^{-i\phi} - \hat{\rho}_{12}e^{i\phi}) \\ &+ \Gamma_{32}\hat{\rho}_{33}, \\ \partial_t\hat{\rho}_{12} =& i\Omega_p\hat{\rho}_{32} - i\Omega_c\hat{\rho}_{13} + i\Omega_w e^{-i\phi}\,(\hat{\rho}_{22} - \hat{\rho}_{11}) \\ &+ i\,(\Delta_p - \Delta_c)\,\hat{\rho}_{12}, \\ \partial_t\hat{\rho}_{13} =& i\Omega_w e^{-i\phi}\hat{\rho}_{23} - i\Omega_c\hat{\rho}_{12} + i\Omega_p(\hat{\rho}_{33} - \hat{\rho}_{11}) \\ &+ \left[i\Delta_p - (\Gamma_{31} + \Gamma_{32})/2\right]\hat{\rho}_{13}, \\ \partial_t\hat{\rho}_{23} =& i\Omega_w e^{i\phi}\hat{\rho}_{13} - i\Omega_p\hat{\rho}_{21} + i\Omega_c\,(\hat{\rho}_{33} - \hat{\rho}_{22}) \\ &+ \left[i\Delta_c - (\Gamma_{31} + \Gamma_{32})/2\right]\hat{\rho}_{23}.\end{aligned} \tag{5}$$

Based on the interaction Hamiltonian in Eq. (4), we can get the equations of the density matrix restricted by the conjugation condition $\hat{\rho}_{ij} = \hat{\rho}^{*}_{ji}$ and the trace condition $\sum_i \hat{\rho}_{ii} = 1$ in Eqs. (5) with $i\,(j) = 1$, 2 and 3. Here $\Gamma_{31}$ denotes the population decay rate from level $|3\rangle \to |1\rangle$ , $\Gamma_{32}$ denotes the population decay rate from level $|3\rangle \to |2\rangle$. Under the steady state $\partial_t\hat{\rho}_{ij} = 0$, we can obtain the numerical solution of $\hat{\rho}_{31}$, which is governed by the probe detuning $\Delta_p$. Then, the steady-state average

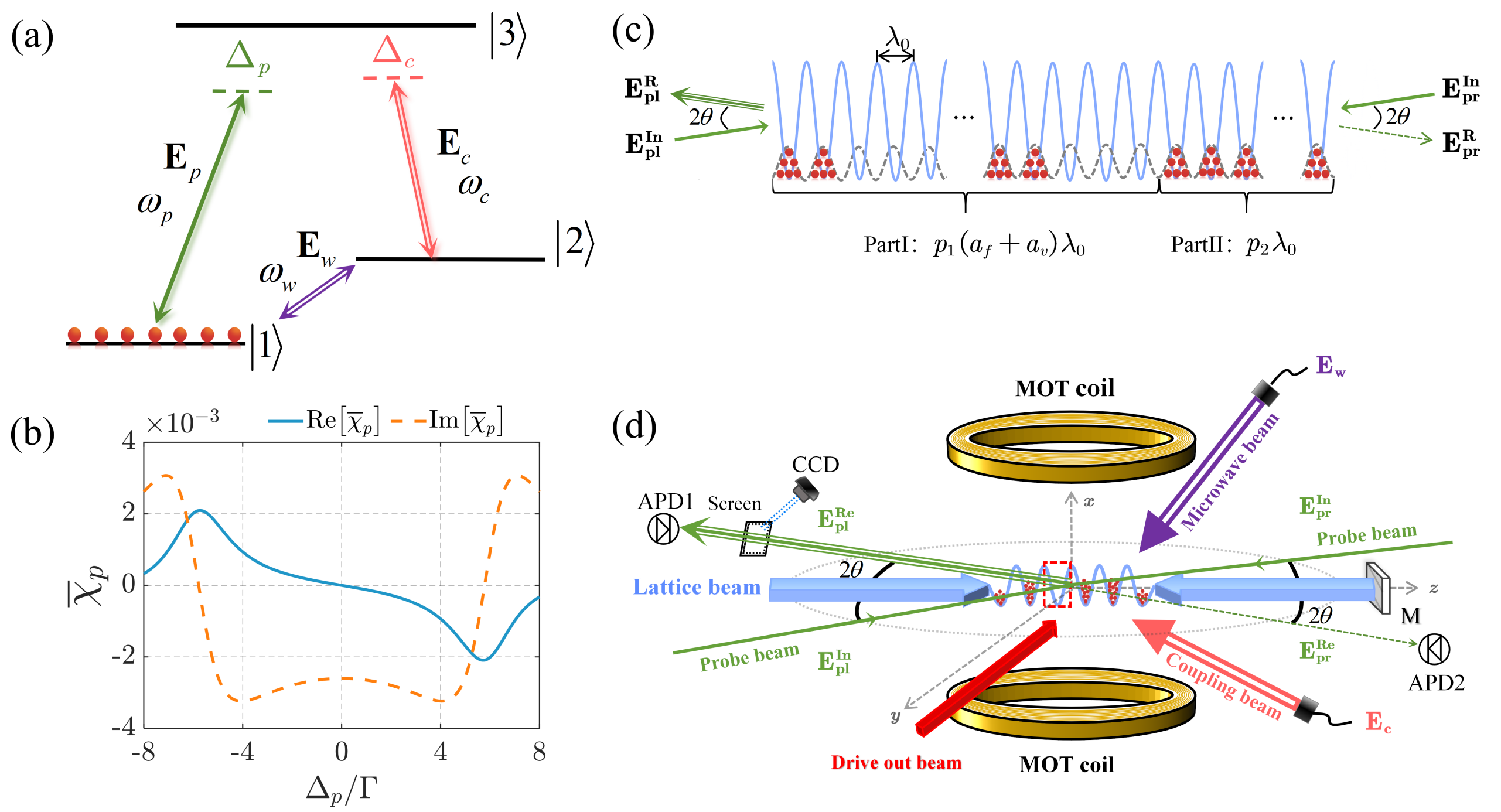

FIG. 1. (a) A coherent gain atom system driven by a coupling field, a probe field and a microwave field. (b) The average probe susceptibility $\overline{\chi}_p$ against the probe detuning $\Delta_p$, with $N_0 = 8.8 \times 10^{10}$ cm$^{-3}$, $\Omega_p = 0.1\Gamma$, $\Omega_c = 6\Gamma$, $\Omega_w = \Gamma$, $\phi = \pi/2$, $\Gamma_{31} = \Gamma$, $\Gamma_{32} = \Gamma$, $\Gamma = 5.75$ MHz. (c) A one-dimensional defect atomic lattice with the width $\lambda_0$ for each period. The probe field is incident from either the left side or the right side denoted by $\mathbf{E}_{\mathbf{pl}}^{\mathbf{In}}$ or $\mathbf{E}_{\mathbf{pr}}^{\mathbf{In}}$ (the relevant reflected fields are denoted by $\mathbf{E}_{\mathbf{pl}}^{\mathbf{Re}}$ and $\mathbf{E}_{\mathbf{pr}}^{\mathbf{Re}}$) with a small incident angle $\theta$ relative to $z$-axis. (d) Schematic of the experimental setup. MOT: magneto-optical trap, APD1: avalanche photodiode used to detect reflected light on the left side, APD2: avalanche photodiode used to detect reflected light on the right side, CCD: charge-coupled device, M: mirror. The parameters: $\lambda_0 = \lambda_{Lat}/2 = 398.4$ nm, $\lambda_{31} = 794.978$ nm and $\eta_0 = 5$. Where $\lambda_{Lat}$ is the wavelength of the retroreflected red-detuned dipole-trap beam (Lattice beam), realizes an atomic Bragg mirror.

probe susceptibility $\overline{\chi}_p$ of each filled cell can be obtained as follows:

$$\overline{\chi}_p = \frac{N_0 |\mathbf{d}_{13}|^2 \hat{\rho}_{31}}{\hbar \varepsilon_0 \Omega_p}, \tag{6}$$

with $N_0$ is the number density of atoms, $d_{13} = 1.0357 \times 10^{-29}$ C $\cdot$ m. Further, we can obtain the average refractive index $\overline{n}_p = \sqrt{1 + \overline{\chi}_p}$. In our system, the effect of the microwave field $\mathbf{E_w}$ enables the system to form a closed-loop, thereby achieving the gain of the probe field $\mathbf{E_p}$. In Fig. 1(b), when the relative phase $\phi = \pi/2$, a broad gain window can be obtained around the resonant frequency. The above discussion is about the average probe susceptibility based on $N_0$. When we consider a large number of cold atoms loaded into a 1D optical lattice, we should envision an atomic density grating of period $\lambda_0$ and assume a Gaussian distribution for $\sigma_z = \lambda_{Lat}/(2\pi\sqrt{\eta_0})$ is the half-width with $\eta_0 = 2U_0/(\kappa_B T)$ related to the capture depth $U_0$ and temperature $T$. For the Bragg condition, the probe field will incident with a little angle $\theta = \arccos\left[\lambda_p/\left(\lambda_{\mathrm{Lat}} - \Delta\lambda_{\mathrm{Lat}}\right)\right]$ along $z$-axis with the geometric Bragg shift $\Delta\lambda_{Lat}$. It should be noted that condition $\lambda_{Lat} > \lambda_{31}$ need to be satified to trap the atoms. Thus, we need to replace the average atomic density $N_0$ in Eq. (6) with the atomic density $N(z)$, where

$$N(z) = \frac{N_0 \lambda_0}{\sigma_z \sqrt{2\pi}} e^{\left[-(z-z_0)^2/2\sigma_z^2\right]}. \tag{7}$$

Then, the Eq. (6) can be rewritten as follows:

$$\chi_p(z) = \frac{N(z) |\mathbf{d_{13}}|^2 \hat{\rho}_{31}}{\hbar \varepsilon_0 \Omega_p}. \tag{8}$$

Further we can get the spatially periodic refractive index $n_p(z) = \sqrt{1 + \chi_p(z)}$.

The reflection and transmission properties of the defective lattice can be characterized by a $2 \times 2$ unimodular transfer matrix. For filled lattice cells, we first divide each cell into 100 thin layers of the same thickness but different density $N(z_s)$ with $s \in \{1, 100\}$, which can be regarded as homogeneous media. The primary transfer matrix of each layer can be written as follows:

$$m(z_s) = \frac{1}{t(z_s)} \begin{bmatrix} t(z_s)^2 - r(z_s)^2 & r(z_s) \\ -r(z_s) & 1 \end{bmatrix}. \tag{9}$$

Then we can obtain the transfer matrix of one filled cell as

$$M_f = \Pi_{s=1}^{100} m(z_s). \tag{10}$$

For a vacuum lattice cell, the transfer matrix can be written as unit for

$$M_v = \frac{1}{t(z)} \begin{bmatrix} t(z)^2 & 0 \\ 0 & 1 \end{bmatrix} = \begin{bmatrix} e^{ik_p \cos\theta\lambda_0} & 0 \\ 0 & e^{-ik_p \cos\theta\lambda_0} \end{bmatrix}, \tag{11}$$

with $k_p = 2\pi/\lambda_p$. Then, the total transfer matrix of the incident light on both sides are represented as follows:

$$M = (M_f)^{p_2} \cdot [(M_v)^{a_v} \cdot (M_f)^{a_f}]^{p_1}. \tag{12}$$

Further, we can calculate the transmission coefficient $t_p$ and the reflection coefficients $r_p^l(r_p^r)$ of the left (right) side of the 1D defective atomic lattice, which are given by

$$t_p = \frac{1}{M(2,2)},\ r_p^l = -\frac{M(2,1)}{M(2,2)},\ r_p^r = \frac{M(1,2)}{M(2,2)}. \tag{13}$$

The transmittance and reflectivities on both sides of the 1D defective atomic lattice can be obtained as follows:

$$T_p = |t_p|^2 = \left|\frac{1}{M(2,2)}\right|^2, \tag{14}$$

$$R_p^l = |r_p^l|^2 = \left|-\frac{M(2,1)}{M(2,2)}\right|^2, \tag{15}$$

$$R_p^r = |r_p^r|^2 = \left|\frac{M(1,2)}{M(2,2)}\right|^2, \tag{16}$$

Similarly, the $S$ matrix relates the outgoing electric field amplitudes $E_{pr}^{out}$ and $E_{pl}^{out}$ to the incoming electric field amplitudes $E_{pl}^{in}$ and $E_{pr}^{in}$ [see Fig. 1(c)], which is defined as follows:

$$\begin{bmatrix} E_{pr}^{out} \\ E_{pl}^{out} \end{bmatrix} = S \begin{bmatrix} E_{pl}^{in} \\ E_{pr}^{in} \end{bmatrix} = \begin{bmatrix} t_p & r_p^r \\ r_p^l & t_p \end{bmatrix} \begin{bmatrix} E_{pl}^{in} \\ E_{pr}^{in} \end{bmatrix}. \tag{17}$$

Then, we can obtain the two eigenvalues of the scattering matrix $S$ as follows:

$$\lambda_S^{\pm} = t_p \pm \sqrt{r_p^l r_p^r}. \tag{18}$$

Here, we use the eigenvalues $\lambda_{S^{-1}}^{\pm}$ of the inverse scattering matrix $S^{-1}$ to verify the existence of lasing namely

$$\lambda_{S^{-1}}^{\pm} = \frac{1}{t_p \pm \sqrt{r_p^l r_p^r}}. \tag{19}$$

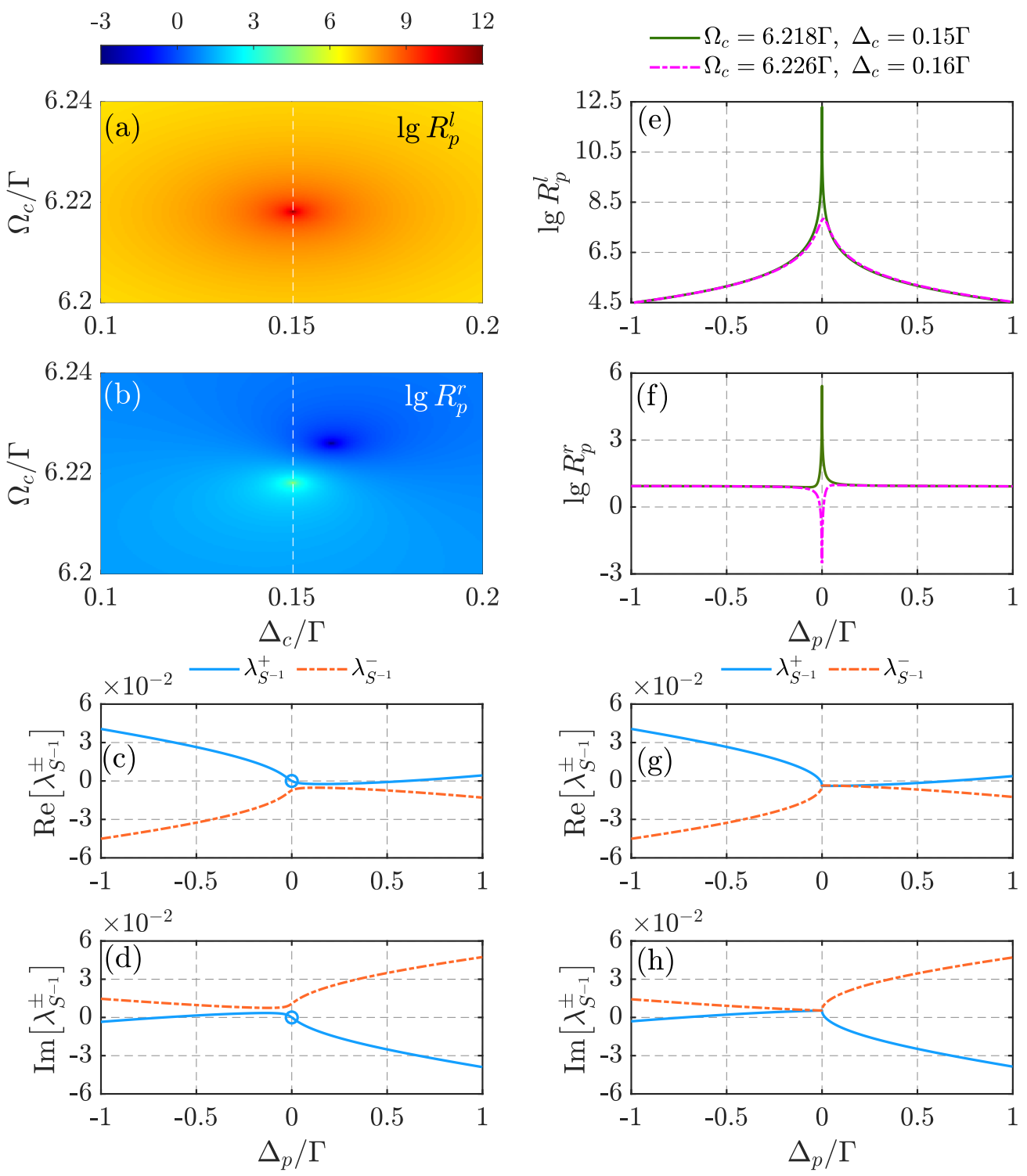

FIG. 2. The logarithm of reflectivities on both sides lg $R_p^l$ and lg $R_p^r$, respectively, against both $\Delta_c$ and $\Omega_c$ in (a) and (b) with $\Delta_p = 0$. The lg $R_p^l$ and lg $R_p^r$ against $\Delta_p$ in (e) and (f). The corresponding eigenvalues Re $[\lambda_{S^{-1}}^{\pm}]$ and Im $[\lambda_{S^{-1}}^{\pm}]$ of $S^{-1}$ against $\Delta_p$ with $\Omega_c = 6.218\Gamma$, $\Delta_c = 0.15\Gamma$ in (c) and (d); $\Omega_c = 6.226\Gamma$, $\Delta_c = 0.16\Gamma$ in (g) and (h). With $a_f = 5$, $a_v = 110$, $p_1 = 20$, $p_2 = 500$, $\Delta\lambda_{Lat} = 0.9$ nm. Other parameters not mentioned are the same as Fig. 1.

## III. NUMERICAL RESULTS AND DISCUSSIONS

First, we need to elaborate in detail on the physical essence of URL in this system. From Eq. (18), it can be clearly observed that when either the left or right reflection coefficient approaches zero ($r_p^r \to 0$ or $r_p^l \to 0$), the two eigenvalues of the scattering matrix tend to degenerate [$\lambda_S^+ \simeq \lambda_S^- \simeq t_p$, leading to the disappearance of $R_p^l$ or $R_p^r$, called non-Hermitian degeneracy (NHD) [50]]. Then, if $\lambda_S^+$ approaches infinity [$\lambda_S^+ \to\infty$, called spectral singularity (SS) [57]], indicates that this eigenvalue is divergent, thereby satisfying the lasing threshold condition. Therefore, when both of the above relationships are satisfied, it corresponds to $\lambda_S^+ \simeq \lambda_S^- \to \infty$, this means a non-Hermitian degeneracy meets a spectral singularity, implying that both eigenvalues and eigenstates tend to degenerate, this is hereafter referred to as non-Hermitian degenerate spectral singularity (NHDSS). This relationship can also be characterized by the eigenvalues of the inverse scattering matrix as: $\lambda_{S^{-1}}^+ \simeq \lambda_{S^{-1}}^- \to 0$, which corresponds to the URL.

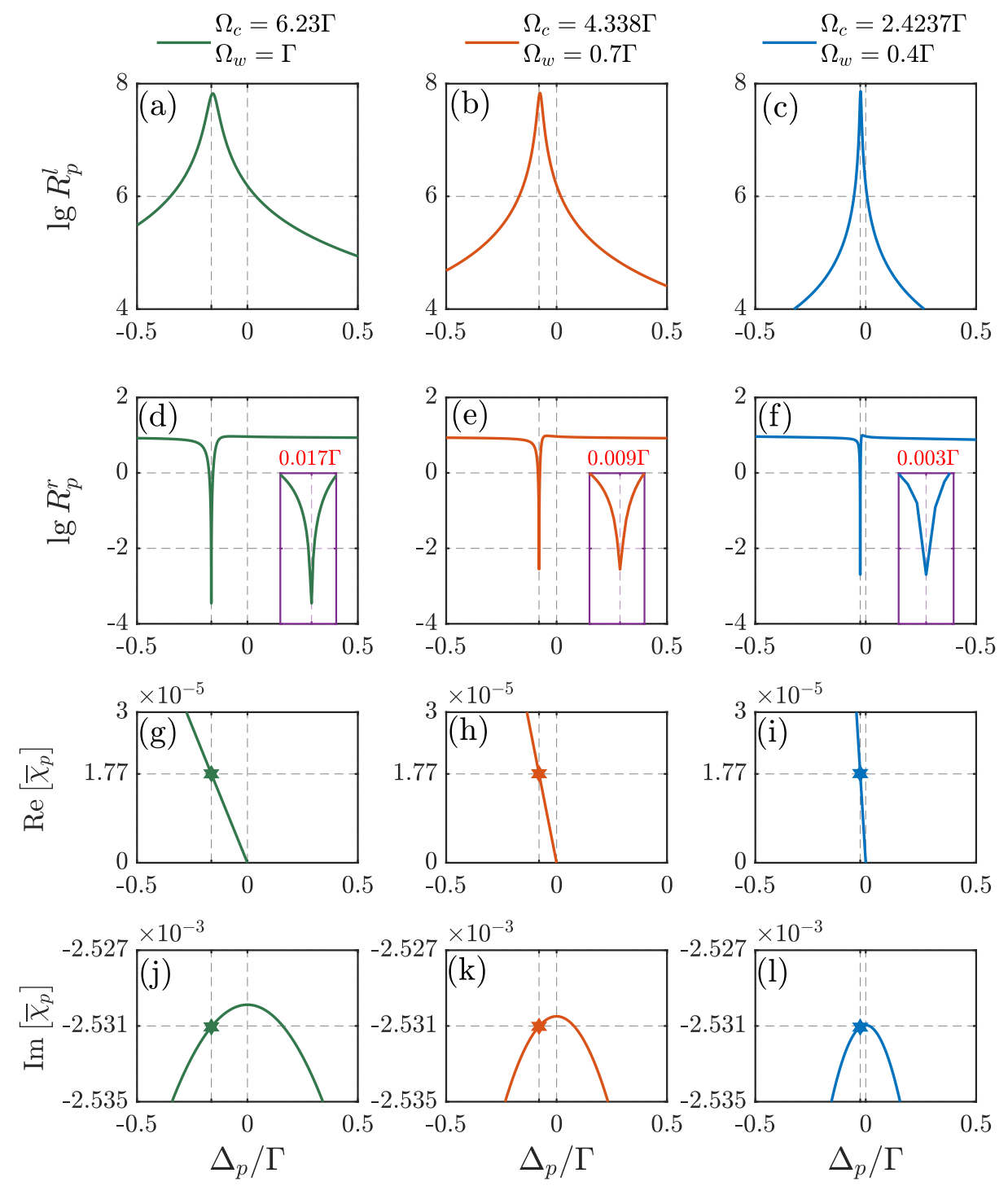

FIG. 3. The $\lg R_p^l$ against $\Delta_p$ in (a) - (c). The $\lg R_p^r$ against $\Delta_p$ in (d) - (f). The real part of the average probe susceptibility $\mathrm{Re}[\overline{\chi}_p]$ against $\Delta_p$ in (g) - (i). The imaginary part of the average probe susceptibility $\mathrm{Im}[\overline{\chi}_p]$ against $\Delta_p$ in (j) - (l). Here, $\Delta_c = 0$, other parameters not mentioned are the same as Fig. 2.

In the following, we adjust the relevant parameters at the probe resonance $\Delta_p = 0$ with the aim of investigating the URL with the relation $\lambda^+_{S^{-1}} \simeq \lambda^-_{S^{-1}} \to 0$. From the pseudo-color maps in Figs. 2(a) and 2(b), which show how $\lg R_p^l$ and $\lg R_p^r$ , respectively, vary with $\Omega_c$ and $\Delta_c$, we observe that at $\Omega_c = 6.218\ \Gamma$, $\Delta_c = 0.15\ \Gamma$, the left and right reflections are non-reciprocal and extremely strong, which can be seen more clearly by the green solid lines in Figs. 2(e) and 2(f). Correspondingly, Figs. 2(c) and 2(d) show the positive eigenvalue of the inverse scattering matrix $\lambda^+_{S^{-1}} \to 0$ (especially, $\lambda^+_{S^{-1}} \to 0$ means that the nonreciprocal reflection lasing (NRL) has been achieved). Intriguingly, the right reflection almost disappears while the left reflection remains very strong with a slight adjustment of $\Omega_c$ and $\Delta_c$ [when $\Omega_c = 6.226\ \Gamma$, $\Delta_c = 0.16\ \Gamma$ in Figs. 2(a) and 2(b)], which can be more clearly observed as the magenta dash-dotted lines in Figs. 2(e) and 2(f). Correspondingly, the two eigenvalues of the inverse scattering matrix $\lambda^+_{S^{-1}} = \lambda^-_{S^{-1}} \to 0$ are exhibited in Figs. 2(g) and 2(h). This implies that the system reaches the lasing threshold when one eigenvalue of the inverse scattering matrix approaches zero. Meanwhile, the system exhibits unidirectionality when the two eigenvalues are nearly equal. At the URL point, we can obtain that the reflected light on the left reaches the lasing threshold condition, while the reflected light on the right meets the unidirectional reflection condition. Furthermore, the lasing intensity of URL is related to the degree of $\lambda^\pm_{S^{-1}} \to 0$ (equivalent to the divergence of the two eigenvalues $\lambda^\pm_S$), while nonreciprocity of the system is related to the degree of the two eigenvalues degeneracy. Thus, $\lambda^\pm_{S^{-1}}$ can be used to characterize and verify the URL and the NRL.

It is straightforward to understand that the extent of $R_p^r \to 0$ determines the degree of $\lambda^+_{S^{-1}} \simeq \lambda^-_{S^{-1}}$. Subsequently, we directly analyze and discuss the points where $R_p^r$ approaches zero and $R_p^l$ is enhanced as the URL, without further elaborating $\lambda^\pm_{S^{-1}}$. To conduct an in-depth analysis of multi-mode URL (i.e., $\lambda^+_{S^{-1}} = \lambda^-_{S^{-1}} \to 0$), we plot the right and left reflectivities and the average susceptibility $\overline{\chi}_p$ by adjusting $\Omega_c$ and $\Omega_w$ in Fig. 3. Notably, we can obtain the narrow band URL with different modes (corresponding to different probe detunings) by modulating $\Omega_c$ and $\Omega_w$, the right reflectivities are consistently constrained below $10^{-2}$ and the left reflectivities remain at $10^8$ [see Figs. 3(a) - 3(f)]. Nevertheless, these modes correspond to the same average susceptibility [Im $[\overline{\chi}_p] \approx -2.531 \times 10^{-3}$ and Re $[\overline{\chi}_p] \approx -1.77 \times 10^{-5}$ see Figs. 3(g) - 3(l)]. This is because only when the probe susceptibility corresponding to the NHDSS point and satisfies the Bragg condition, it can perfectly match the URL. It can be seen that under a specific lattice structure, the position and width of the URL can be manipulated by adjusting the external optical field. More notably, the linewidth of the URL can be as low as 20 kHz ($0.003\Gamma$) with $\lg R_p^r \leqslant 0$. Therefore, the URL holds irreplaceable applications in precision measurement, quantum information, optical communication and other fields.

The emergence of nonreciprocity (or even unidirectionality) from both sides of reflections is a consequence of the introduction of vacant lattice cells $a_v$, which breaks the spatial symmetry of the probe susceptibility. Fundamentally, this phenomenon arises from the fact that the probe phase is modulated by the vacuum lattice cells during propagation, which is determined by the modifications to the Bragg condition [46]. Now, we first investigate the logarithm of reflectivities $R_p^{l,r}$ *v.s.* vacant lattice cells $a_v$ under different geometric Bragg shift $\Delta\lambda_{Lat}$ in Figs. 4(a) - 4(c). The reflectivities on both sides are reciprocal at $a_v = 0$, As $a_v$ increases, they begin to oscillate periodically and eventually become reciprocal after 20 periods. Notably, these periods are equal to $p_1$ and are unaffected by $\Delta\lambda_{Lat}$. Because the probe will accumulate a certain phase after passing through vacant lattice cells determined by $a_v$, the reflections become reciprocal when the phase accumulation reaches $2\mathbb{N}\pi$, where $\mathbb{N}$ is a non-negative integer. Under the representation of the vacant cell, the condition for the system to return to reciprocity is as follow:

$$a_v \cdot \Delta\lambda_{Lat} = \mathbb{N}\left(\lambda_{Lat} - \Delta\lambda_{Lat}\right). \tag{20}$$

This relationship, which is equivalent to the condition

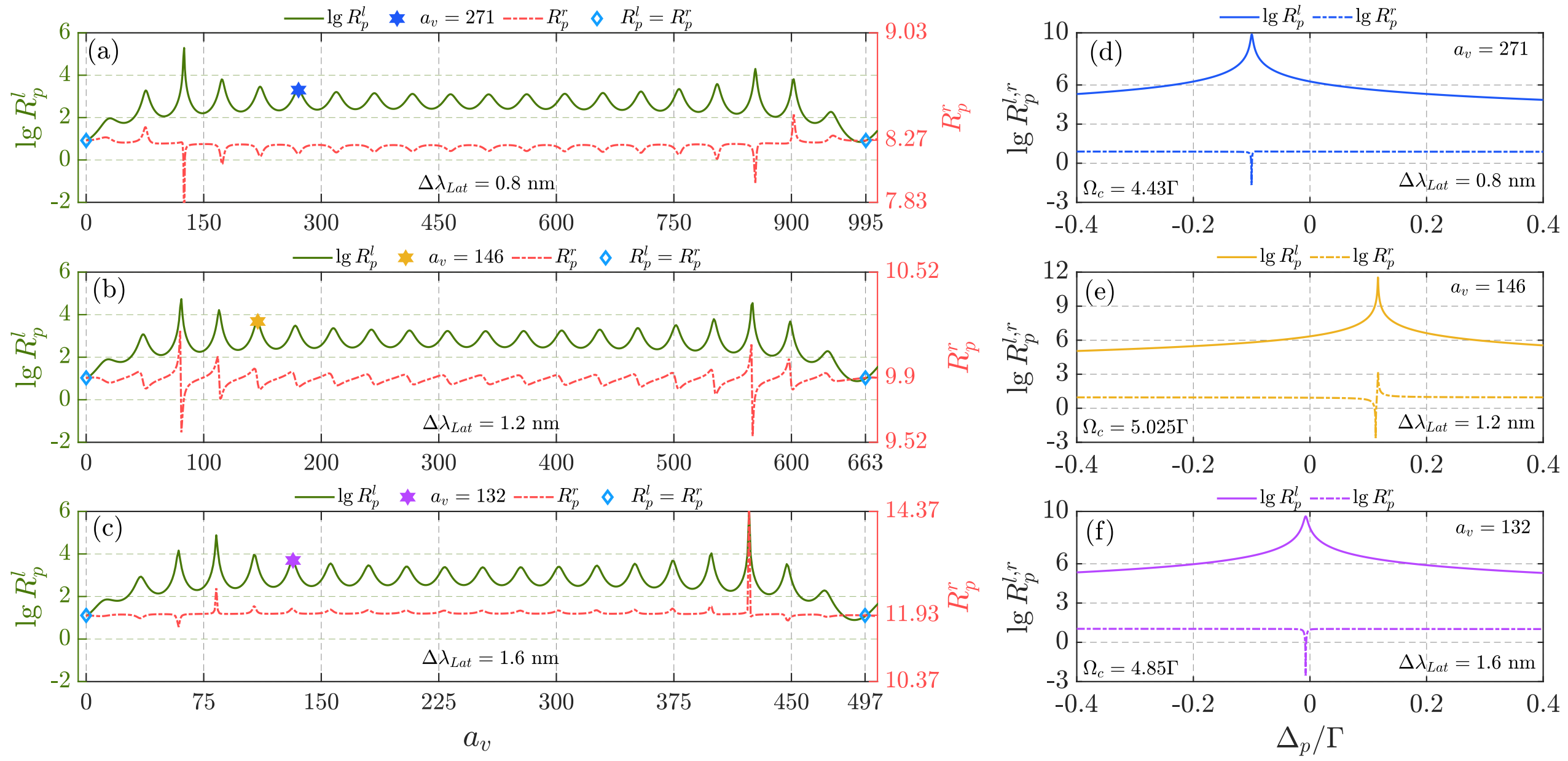

FIG. 4. The $\lg R_p^l$ and $R_p^r$ against the vacant lattice cells $a_v$ in (a) - (c) under different geometric Bragg shift $\Delta\lambda_{Lat}$ with $\Delta_p = 0$, $\Omega_c = 6.226\Gamma$ and $\Delta_c = 0.16\Gamma$. The $\lg R_p^{l,r}$ against $\Delta_p$ in (d) - (f) with $\Delta_c = 0$. Other parameters not mentioned are the same as Fig. 2.

$k_p \cos\theta\lambda_0 \cdot a_v = 2\mathbb{N}_0\pi$ with $\mathbb{N}_0$ is an non-negative integer ($\mathbb{N}_0 \neq \mathbb{N}$), can be obtained through Eq. (11). Notably, under the parameters in Figs. 4(a) - 4(c), the URL could not be achieved consistently. Though the left reflectivity has reached the intensity of lasing output, the right reflectivity is also amplified, and the positions of the peaks and valleys of the left and right reflections also depend on $a_v$. However, via proper parameter adjustment, each oscillation period with satisfying the Bragg conditions can form the URL at different probe frequencies (excluding the first and last oscillations), for details, see Appendix B. Next, we randomly select a value of $a_v$ from Figs. 4(a), 4(b), and 4(c), respectively. These values correspond to the peak of the left reflection, while the right reflection corresponds to the peak, peak valley mutation, and valley. With these three values of $a_v$, we then adjust $\Omega_c$ to check the left and right reflectivities $\lg R_p^{l,r}$ *v.s.* $\Delta_p$ in Figs. 4(d) - 4(f). It can be observed that when the reflectivities on both sides vary with the number of vacant lattice cells (either both increasing/decreasing or one increasing while the other decreasing), a relatively ideal URL can be modulated, as illustrated in Figs. 4(d) and 4(f). In contrast, at the position of peak-valley mutations between the left and right reflection, the NHD and SS cannot be satisfied simultaneously, corresponding to the NRL depicted in Fig. 4(e).

## IV. EXPERIMENTAL FEASIBILITY

According to experimental Refs. [44, 47, 48], we design the corresponding experimental setup in Fig. 1(d). The $^{87}$Rb atoms are first cooled via a magneto-optical trap (MOT), then trapped in an optical lattice formed by a retroreflected beam to construct a perfect atomic lattice. Subsequently, atoms in pre-designated filled lattice cells are thermally expelled using a drive out beam, creating vacant lattice cells and thus fabricating a defective atomic lattice. The coupling and microwave beams cover the entire defective atomic lattice, and the probe beam is incident from the left or right side at a small angle $\theta$ with respect to the $z$-axis. The distributed feedback (DFB) lasing (probe beam is incident from the left side, corresponds to NHDSS and satisfies the Bragg condition) can be detected by APD1 (avalanche photodiode). The beam cross-section is observed using a charge-coupled device (CCD) camera and a movable light screen S. In contrast, the probe beam incident from the right side exhibits reflectionless behavior, thus the APD2 detect only an extremely weak signal or no signal at all.

## V. CONCLUSIONS

In this letter, we achieve controllable URL by leveraging gain atomic systems induced by a microwave field, symmetry breaking caused by a defective atomic lattice, and distributed feedback. Naturally, this requires that the eigenvalues $\lambda^{\pm}_{S^{-1}}$ of the inverse scattering matrix $S^{-1}$ exhibit NHD and SS ($\lambda^+_{S^{-1}} \simeq \lambda^-_{S^{-1}} \to 0$). Moreover, the perfect URL corresponds to a left reflectivity as high as $10^8$ and a right reflectivity as low as $10^{-2}$ along with an adjustable linewidth. Notably, when the lattice structure is fixed, the multimode URL emerges as the average susceptibility $\overline{\chi}_p$ is maintained at its deterministic value by tuning the parameters of the additional optical fields, this value of $\overline{\chi}_p$ strictly satisfies the Bragg condition, a function of lattice parameters and the probe susceptibility. In other words, the URL not only requires correspondence to NHDSS point but also necessitates fulfillment of the Bragg condition, this demands the combined modulation of the lattice structure and external fields, and we have provided precise, experimentally feasible parameters in this work. These findings offer potential pathways for the development of nonreciprocal optical circuits and high-performance all-optical controlled unidirectional devices in integrated photonic circuits [63, 64].

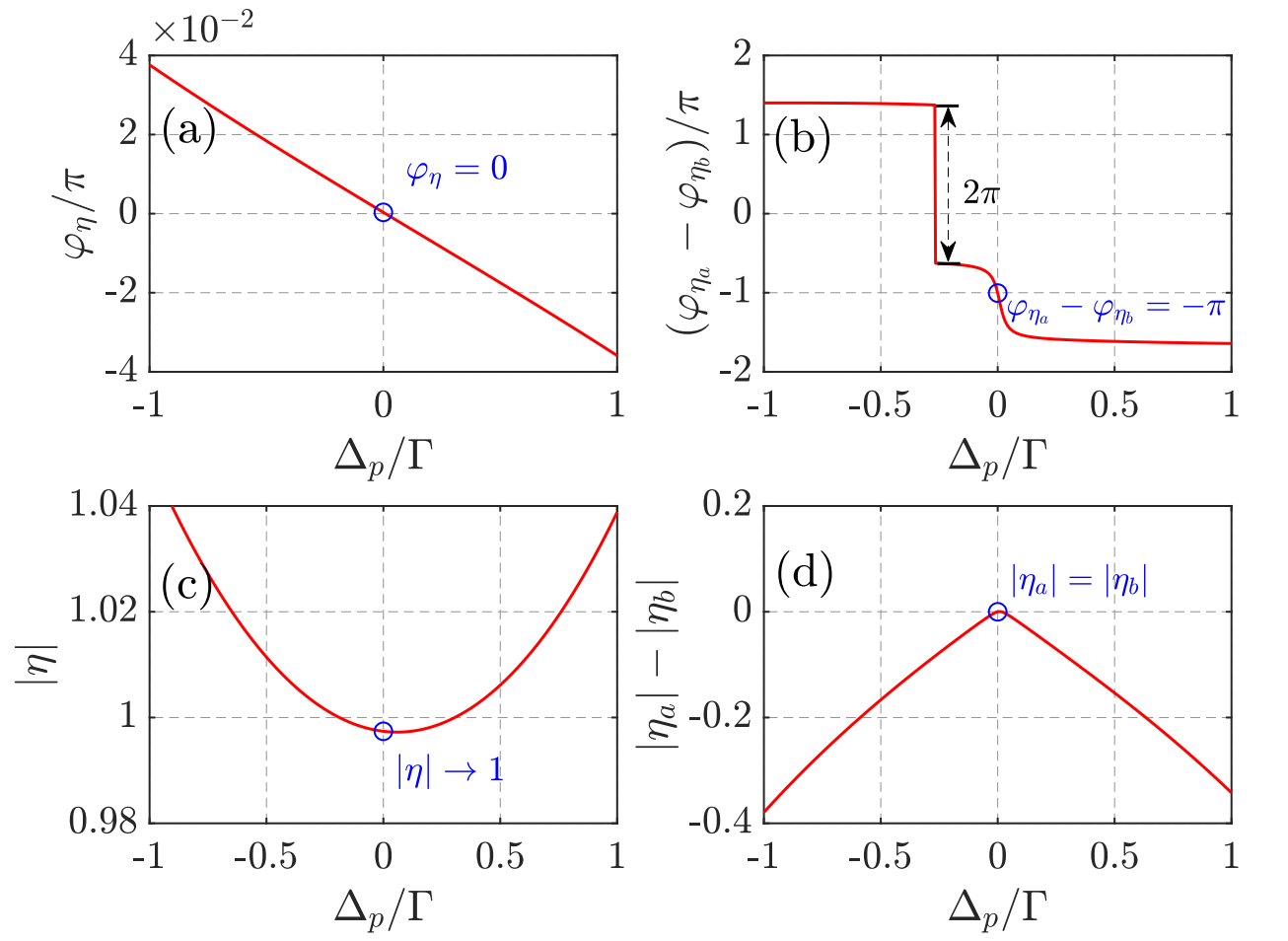

FIG. 5. The amplitude $|\eta|$ and phase $\varphi_\eta$ of $\eta$ against $\Delta_p$ in (a) and (c). The difference in amplitude $|\eta_a| - |\eta_b|$ and phase $\varphi_{\eta_a} - \varphi_{\eta_b}$ against $\Delta_p$ in (b) and (d). The parameters are the same as the magenta dash-dotted lines in Figs. 2(e) and 2(f).

## VI. ACKNOWLEDGMENTS

This work is supported by the National Natural Science Foundation of China (Grant Nos. 12204137, 12564048, 12564047, 11874004, 11204019) and Hainan Provincial Graduate Innovative Research Project (Grant No. Qhys2024-402).

## Appendix A: THE PHYSICAL ESSENCE OF URL

From Eq. (12), we can split the total transport matrix $M$ of the 1D defective atomic lattice into two parts, which respectively represent the transfer matrix of part I ($M_{\rm I}$) and part II ($M_{\rm II}$) as follows:

$$M_{\rm I} = \left[(M_v)^{a_v} \cdot (M_f)^{a_f}\right]^{p_1},\ M_{\rm II} = (M_f)^{p_2}. \tag{A.1}$$

Through simple matrix operations and substitutions, we can obtain the following relationship

$$\begin{aligned} M(1,1) &= M_{\rm I}(1,1) \cdot M_{\rm II}(1,1) + M_{\rm I}(2,1) \cdot M_{\rm II}(1,2), \\ M(1,2) &= M_{\rm I}(1,2) \cdot M_{\rm II}(1,1) + M_{\rm I}(2,2) \cdot M_{\rm II}(1,2), \\ M(2,1) &= M_{\rm I}(1,1) \cdot M_{\rm II}(2,1) + M_{\rm I}(2,1) \cdot M_{\rm II}(2,2), \\ M(2,2) &= M_{\rm I}(1,2) \cdot M_{\rm II}(2,1) + M_{\rm I}(2,2) \cdot M_{\rm II}(2,2). \end{aligned} \tag{A.2}$$

Here, The expressions of the transmission and reflection coefficients on both sides of the two parts are as follows:

$$\begin{aligned} t_{\rm I} &= \frac{1}{M_{\rm I}(2,2)},\ r^l_{\rm I} = -\frac{M_{\rm I}(2,1)}{M_{\rm I}(2,2)},\ r^r_{\rm I} = \frac{M_{\rm I}(1,2)}{M_{\rm I}(2,2)}, \\ t_{\rm II} &= \frac{1}{M_{\rm II}(2,2)},\ r^l_{\rm II} = -\frac{M_{\rm II}(2,1)}{M_{\rm II}(2,2)}, r^r_{\rm II} = \frac{M_{\rm II}(1,2)}{M_{\rm II}(2,2)}. \end{aligned} \tag{A.3}$$

Here, $t_{\rm I(II)}$, $r^l_{\rm I(II)}$ and $r^r_{\rm I(II)}$ are the transmission coefficients, the left and right reflection coefficients of part I (part II), respectively. Furthermore, based on Eq. (13), we can represent each matrix element by transmission and reflection coefficients on both sides as follows:

$$\begin{aligned} M_{\rm I}(1,1) &= t_{\rm I} - \frac{r^l_{\rm I} r^r_{\rm I}}{t_{\rm I}},\ M_{\rm I}(1,2) = \frac{r^r_{\rm I}}{t_{\rm I}}, \\ M_{\rm I}(2,1) &= -\frac{r^l_{\rm I}}{t_{\rm I}},\ M_{\rm I}(2,2) = \frac{1}{t_{\rm I}}, \\ M_{\rm II}(1,1) &= t_{\rm II} - \frac{r^l_{\rm II} r^r_{\rm II}}{t_{\rm II}},\ M_{\rm II}(1,2) = \frac{r^r_{\rm II}}{t_{\rm II}}, \\ M_{\rm II}(2,1) &= -\frac{r^l_{\rm II}}{t_{\rm II}},\ M_{\rm II}(2,2) = \frac{1}{t_{\rm II}}. \end{aligned} \tag{A.4}$$

Therefore, the matrix elements $M(2,2)$ and $M(1,2)$ of the total transfer matrix can be replaced by

$$M(2,2) = \frac{1 - r^r_{\rm I} r^l_{\rm II}}{t_{\rm I} t_{\rm II}} = \frac{1-\eta}{t_{\rm I} t_{\rm II}}, \tag{A.5}$$

$$M(1,2) = \frac{t_{\rm II} r^r_{\rm I} t_{\rm II} + \left(1 - r^r_{\rm I} r^l_{\rm II}\right) r^r_{\rm II}}{t_{\rm I} t_{\rm II}} = \frac{\eta_a + \eta_b}{t_{\rm I} t_{\rm II}}. \tag{A.6}$$

Obviously, as $(1-\eta) \to 0$ we have $M(2,2) \to 0$, where the system satisfies the threshold condition, accompanied by the generation of the transmission and reflection lasing. To clarify the underlying physics, we present the complex-valued expressions as follows: $\eta = r^r_{\rm I} r^l_{\rm II} = |\eta| \cdot e^{i\varphi_\eta}$, $\eta_a = t^l_{\rm II} r^r_{\rm I} t^r_{\rm II} = |\eta_a| \cdot e^{i\varphi_{\eta_a}}$ and $\eta_b = \left(1 - r^r_{\rm I} r^l_{\rm II}\right) r^r_{\rm II} = |\eta_b| \cdot e^{i\varphi_{\eta_b}}$. Therefore, $|\eta| \to 1$ corresponding to $\varphi_\eta \to 0$ can be adopted as a criterion to confirm the lasing, which exerts a decisive influence on reflectance at both interfaces. From Eq. (16), the reflectivity on the right side of the 1D defective atomic lattice is determined by $M(1,2)$ and $M(2,2)$. At the

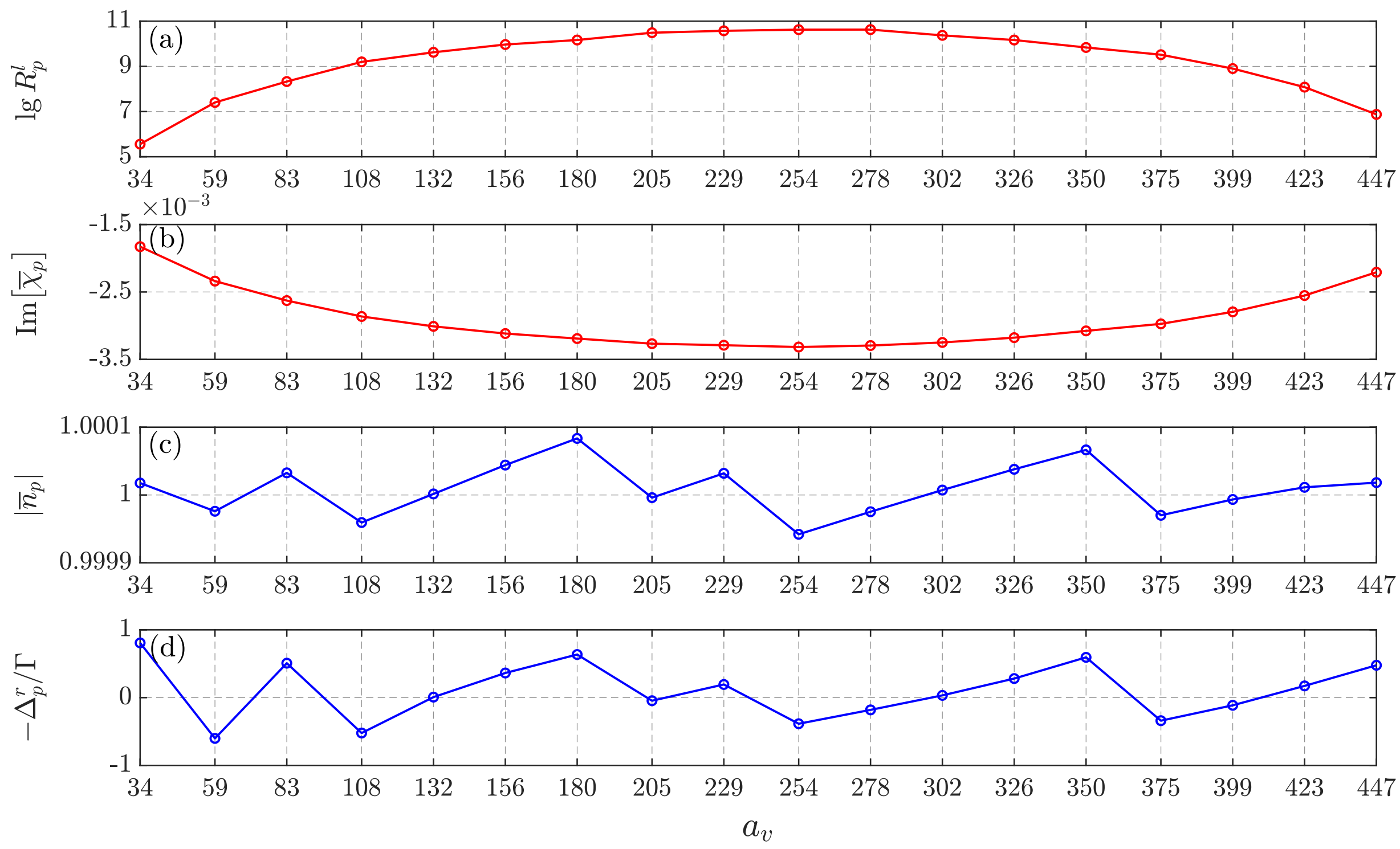

FIG. 6. Under different vacant cells, the logarithm of reflectivity on the left side $\lg R_p^l$, the imaginary part of the average probe susceptibility $\mathrm{Im}[\overline{\chi}_p]$, the magnitude of average refractive index $|\overline{n}_p|$, and the probe detuning frequency $-\Delta_p^r/\Gamma$ corresponding to the right reflectivity NHDSS points against $a_v$ in (a)-(d), respectively, with $\Delta\lambda_{Lat} = 1.6$ nm and $\Delta_c = 0$. Expect for $\Omega_c$, other parameters not mentioned are the same as Fig. 2.

URL point, the right reflection vanishes as both $M(1,2)$ and $M(2,2)$ approach zero simultaneously. Under the condition of $M(1,2) \to 0$, we derive $\eta_a - \eta_b = 0$ corresponding to $|\eta_a| = |\eta_b|$ and $\varphi_{\eta_a} - \varphi_{\eta_b} = (2N_\eta + 1)\pi$ with $N_\eta$ is an integer.

To elucidate the physical essence of URL, we plot the amplitude and phase of $\eta$, as well as the amplitude difference $|\eta_a| - |\eta_b|$ and phase difference $\varphi_{\eta_a} - \varphi_{\eta_b}$ against $\Delta_p$ in Fig. 5. As clearly illustrated, at resonance [corresponding to the URL point, marked by magenta dash-dotted lines in Figs. 2(e) and 2(f)], $|\eta| \to 1$ and $\varphi_\eta \to 0$ [see Figs. 5(a) and 5(c)], accompanied by $|\eta_a| = |\eta_b|$ and $\varphi_{\eta_a} - \varphi_{\eta_b} = -\pi$ [see Figs. 5(b) and 5(d)]. This implies $M(2,2) \to 0$ and $M(1,2) \to 0$, which leads to the enhancement of the left reflection ($R_p^l \to \infty$) and the suppression of the right reflection ($R_p^r \to 0$). According to the definition of $\eta_a = t_{\mathrm{II}}^r r_{\mathrm{I}}^r t_{\mathrm{II}}^l$, the propagation process represented by $\eta_a$ is as follows: the probe that incident from the right side of the defective atomic lattice transmits through part II ($t_{\mathrm{II}}^r$), undergoes reflection at part I ($r_{\mathrm{I}}^r$) and then transmits through part II again ($t_{\mathrm{II}}^l$). Concurrently, this portion $\eta_a$ of the probe destructively interferes with the portion of the probe reflected by part II ($r_{\mathrm{II}}^r$), provided that both the amplitude condition $|\eta_a| = |\eta_b|$ and the phase difference $\varphi_{\eta_a} - \varphi_{\eta_b} = -\pi$ are met. This constitutes the fundamental physical mechanism underlying the elimination of right reflection. It is worth noting that the $2\pi$ phase jump observed in Fig. 5 (b) holds no special physical significance, as it is equivalent to a phase change of nearly zero.

## Appendix B: TUNABLE URL

In our system, the lattice structure of the URL is not unique. To verify whether every nonreciprocal peak can be modulated into a URL and explore its physical essence, we plot $\lg R_p^l$, $\mathrm{Im}[\overline{\chi}_p]$, $|\overline{n}_p|$ and $-\Delta_p^r/\Gamma$ against $a_v$ in Fig. 6. Notably, URLs have been successfully achieved at the 18 peak positions of left reflectivity (excluding the first and last oscillations) by adjusting $\Omega_c$ to satisfy NHDSS with $\Delta\lambda_{Lat} = 1.6$ nm. Subsequently, we found that the intensity of the URL is negatively correlated with the imaginary part of the average susceptibility $\mathrm{Im}[\overline{\chi}_p]$, i.e., proportional to the gain amplitude [see Figs. 6(a) and 6(b)]. As $a_v$ increases, the gain amplitude of the susceptibility that satisfying NHDSS first increases and then decreases. Furthermore, when $a_v \in(205, 302)$ the intensity of left reflection lasing reaches extreme values. For instance, $a_v = 229$, $a_v = 254$ and $a_v = 278$, we can obtain three URLs with $\lg R_p^l > 10.5$. It is worth emphasizing that the frequencies of these extremely strong URLs are tunable between -Γ and +Γ, as shown in Fig. 6(d). Interestingly, the probe detuning $-\Delta_p^r/\Gamma$ corre-sponding to the URL is positively correlated with the average refractive index $|\overline{n}_p|$ [see Fig. 6(d)]. This indicates that the intensity of the URLs are directly determined by the imaginary part of the average susceptibility $\mathrm{Im}[\overline{\chi}_p]$ and the frequencies of the URLs are governed by the average refractive index $|\overline{n}_p|$. At these points, the susceptibility must not only ensure $\lambda^{\pm}_{S^{-1}}$ satisfies both NHD and SS but also meet the Bragg condition [46]. A comparison of Fig. 6(c) and 6(d) reveals that the probe detuning of the URL point is modulated by $|\overline{n}_p|$ and varies synchronously with it. Correspondingly, the probe frequency decreases (increases) linearly, enabling the modulation of Bragg condition-satisfying URLs from blue shift to red shift (or red shift to blue shift) under different $a_v$.

[1] K.-W. Huang, X. Wang, Q.-Y. Qiu, L. Wu, and H. Xiong, Nonreciprocal phonon laser in an asymmetric cavity with an atomic ensemble, Chin. Phys. Lett. **40**, 104201 (2023).

[2] X.-B. Chen, L. Fan, R. Zhang, and C. Cao, Nonreciprocal phonon laser in a rotating cavity optomechanical system with an atomic ensemble, Laser Phys. Lett. **22**, 055204 (2025).

[3] Z.-Y. Wang, X.-W. He, X. Han, H.-F. Wang, and S. Zhang, Nonreciprocal pt-symmetric magnon laser in spinning cavity optomagnonics, Opt. Express **32**, 4987 (2024).

[4] Z.-Y. Wang, J. Qian, Y.-P. Wang, J. Li, and J. You, Realization of the unidirectional amplification in a cavity magnonic system, Appl. Phys. Lett. **123** (2023).

[5] R.-T. Sun, M.-Y. Peng, T.-X. Lu, J. Wang, Q. Zhang, Y.-F. Jiao, and H. Jing, Multicolor nonreciprocal optical amplifier with spinning active optomechanics, Phys. Rev. A **109**, 023520 (2024).

[6] E. Galiffi, P. Huidobro, and J. B. Pendry, Broadband nonreciprocal amplification in luminal metamaterials, Phys. Rev. Lett. **123**, 206101 (2019).

[7] T.-T. Zhang, W.-P. Zhou, Z.-X. Li, Y.-T. Tang, F. Xu, H.-H. Wu, H. Zhang, J.-S. Tang, Y.-P. Ruan, and K.-Y. Xia, Reversible optical isolators and quasi-circulators using a magneto-optical fabry–pérot cavity, Chin. Phys. Lett. **41**, 044205 (2024).

[8] Y.-Q. Hu, S.-C. Zhang, Y.-H. Qi, G.-W. Lin, Y.-P. Niu, and S.-Q. Gong, Multiwavelength magnetic-free optical isolator by optical pumping in warm atoms, Phys. Rev. Appl. **12**, 054004 (2019).

[9] E.-Z. Li, D.-S. Ding, Y.-C. Yu, M.-X. Dong, L. Zeng, W.-H. Zhang, Y.-H. Ye, H.-Z. Wu, Z.-H. Zhu, W. Gao, G.-C. Guo, and B.-S. Shi, Experimental demonstration of cavity-free optical isolators and optical circulators, Phys. Rev. Research **2**, 033517 (2020).

[10] N. O. Antoniadis, N. Tomm, T. Jakubczyk, R. Schott, S. R. Valentin, A. D. Wieck, A. Ludwig, R. J. Warburton, and A. Javadi, A chiral one-dimensional atom using a quantum dot in an open microcavity, npj Quantum Inf. **8**, 27 (2022).

[11] L.-N. Song, Q. Zheng, X.-W. Xu, C. Jiang, and Y. Li, Optimal unidirectional amplification induced by optical gain in optomechanical systems, Phys. Rev. A **100**, 043835 (2019).

[12] L. Mercier de Lépinay, E. Damskägg, C. F. Ockeloen-Korppi, and M. A. Sillanpää, Realization of directional amplification in a microwave optomechanical device, Phys. Rev. Appl. **11**, 034027 (2019).

[13] N. T. Otterstrom, E. A. Kittlaus, S. Gertler, R. O. Behunin, A. L. Lentine, and P. T. Rakich, Resonantly enhanced nonreciprocal silicon brillouin amplifier, Optica **6**, 1117 (2019).

[14] E. A. Kittlaus, H. Shin, and P. T. Rakich, Large brillouin amplification in silicon, Nat. Photonics **10**, 463 (2016).

[15] A. Muñoz de las Heras and I. Carusotto, Optical isolators based on nonreciprocal four-wave mixing, Phys. Rev. A **106**, 063523 (2022).

[16] Y. Xu, J.-Y. Liu, W. Liu, and Y.-F. Xiao, Nonreciprocal phonon laser in a spinning microwave magnomechanical system, Phys. Rev. A **103**, 053501 (2021).

[17] Y. Jiang, S. Maayani, T. Carmon, F. Nori, and H. Jing, Nonreciprocal phonon laser, Phys. Rev. Appl. **10**, 064037 (2018).

[18] T.-X. Lu, Y. Wang, K. Xia, X. Xiao, L.-M. Kuang, and H. Jing, Quantum squeezing induced nonreciprocal phonon laser, Sci. China Phys. Mech. Astron. **67**, 1 (2024).

[19] H. Xu, L. Jiang, A. A. Clerk, and J. G. E. Harris, Nonreciprocal control and cooling of phonon modes in an optomechanical system, Nature **568**, 65 (2019).

[20] A. Seif, W. DeGottardi, K. Esfarjani, and M. Hafezi, Thermal management and non-reciprocal control of phonon flow via optomechanics, Nat. Commun. **9**, 1207 (2018).

[21] Q.-S. Wang, Z.-L. Zhou, D.-M. Liu, H. Ding, M. Gu, and Y. Li, Acoustic topological beam nonreciprocity via the rotational doppler effect, Sci. Adv. **8**, eabq4451 (2022).

[22] S. A. Cummer, Selecting the direction of sound transmission, Science **343**, 495 (2014).

[23] Y.-J. Xu and J. Song, Nonreciprocal magnon laser, Opt. Lett **46**, 5276 (2021).

[24] K.-W. Huang, Y. Wu, and L.-G. Si, Parametric-amplification-induced nonreciprocal magnon laser, Opt. Lett **47**, 3311 (2022).

[25] X.-W. He, Z.-Y. Wang, X. Han, S. Zhang, and H.-F. Wang, Parametrically amplified nonreciprocal magnon laser in a hybrid cavity optomagnonical system, Opt. Express **31**, 43506 (2023).

[26] X.-W. He, Z.-Y. Wang, X. Han, H.-F. Wang, and S. Zhang, Nonreciprocal magnon laser in a spinning cavity optomagnonic system, Opt. Lett **50**, 499 (2025).

[27] Z.-Y. Wang, X.-W. He, X. Han, H.-F. Wang, and S. Zhang, Nonreciprocal pt-symmetric magnon laser in spinning cavity optomagnonics, Opt. Express **32**, 4987 (2024).

[28] B. Wang, X. Jia, X.-H. Lu, and H. Xiong, Pt-symmetric magnon laser in cavity optomagnonics, Phys. Rev. A **105**, 053705 (2022).

[29] A. Muñoz de las Heras and I. Carusotto, Unidirectional lasing in nonlinear taiji microring resonators, Phys. Rev. A **104**, 043501 (2021).

[30] N. T. Otterstrom, E. A. Kittlaus, S. Gertler, R. O. Behunin, A. L. Lentine, and P. T. Rakich, Resonantly enhanced nonreciprocal silicon brillouin amplifier, Optica

**6**, 1117 (2019).

[31] L.-N. Song, Z.-H. Wang, and Y. Li, Enhancing optical nonreciprocity by an atomic ensemble in two coupled cavities, Opt. Commun. **415**, 39 (2018).

[32] B. Abdo, K. Sliwa, S. Shankar, M. Hatridge, L. Frunzio, R. Schoelkopf, and M. Devoret, Josephson directional amplifier for quantum measurement of superconducting circuits, Phys. Rev. Lett. **112**, 167701 (2014).

[33] A. Metelmann and A. A. Clerk, Nonreciprocal photon transmission and amplification via reservoir engineering, Phys. Rev. X **5**, 021025 (2015).

[34] A. Hakimi, K. Rouhi, T. G. Rappoport, M. G. Silveirinha, and F. Capolino, Chiral terahertz lasing with berry-curvature dipoles, Phys. Rev. Appl. **22**, L041003 (2024).

[35] T. T. Koutserimpas and R. Fleury, Nonreciprocal gain in non-hermitian time-floquet systems, Phys. Rev. Lett. **120**, 087401 (2018).

[36] W. Guerin, F. Michaud, and R. Kaiser, Mechanisms for lasing with cold atoms as the gain medium, Phys. Rev. Lett. **101**, 093002 (2008).

[37] J. C. d. A. Carvalho, A. D. M. de Lima, and J. W. R. Tabosa, Lasing without inversion based on magnetically assisted gain in coherently prepared cold atoms, Phys. Rev. A **105**, 023706 (2022).

[38] J.-H. Wu, M. Artoni, and G. C. La Rocca, Two-color lasing in cold atoms, Phys. Rev. A **88**, 043823 (2013).

[39] G. W. Harmon, J. T. Reilly, M. J. Holland, and S. B. Jäger, Mean-field floquet theory for a three-level cold-atom laser, Phys. Rev. A **106**, 013706 (2022).

[40] B. Megyeri, G. Harvie, A. Lampis, and J. Goldwin, Directional bistability and nonreciprocal lasing with cold atoms in a ring cavity, Phys. Rev. Lett. **121**, 163603 (2018).

[41] A. Schilke, C. Zimmermann, P. W. Courteille, and W. Guerin, Photonic band gaps in one-dimensionally ordered cold atomic vapors, Phys. Rev. Lett. **106**, 223903 (2011).

[42] H. Yang, L. Yang, X.-C. Wang, C.-L. Cui, Y. Zhang, and J.-H. Wu, Dynamically controlled two-color photonic band gaps via balanced four-wave mixing in one-dimensional cold atomic lattices, Phys. Rev. A **88**, 063832 (2013).

[43] H. Yang, T.-G. Zhang, Y. Zhang, and J.-H. Wu, Dynamically tunable three-color reflections immune to disorder in optical lattices with trapped cold $^{87}$Rb atoms, Phys. Rev. A **101**, 053856 (2020).

[44] A. Schilke, C. Zimmermann, P. W. Courteille, and W. Guerin, Optical parametric oscillation with distributed feedback in cold atoms, Nat. Photonics **6**, 101 (2012).

[45] T.-M. Li, M.-H. Wang, C.-P. Yin, J.-H. Wu, and H. Yang, Dynamic manipulation of three-color light reflection in a defective atomic lattice, Opt. Express **29**, 31767 (2021).

[46] T.-M. Li, H. Yang, M.-H. Wang, C.-P. Yin, T.-G. Zhang, and Y. Zhang, Unidirectional photonic reflector using a defective atomic lattice, Phys. Rev. Research **6**, 023122 (2024).

[47] A. Schilke, C. Zimmermann, P. W. Courteille, and W. Guerin, Photonic band gaps in one-dimensionally ordered cold atomic vapors, Phys. Rev. Lett. **106**, 223903 (2011).

[48] A. Schilke, C. Zimmermann, and W. Guerin, Photonic properties of one-dimensionally-ordered cold atomic vapors under conditions of electromagnetically induced transparency, Phys. Rev. A **86**, 023809 (2012).

[49] Q.-Y. Xu, G.-R. Li, Y.-T. Zheng, D. Yan, H.-X. Zhang, T.-G. Zhang, and H. Yang, Broadband unidirectional reflection amplification in a one-dimensional defective atomic lattice, Phys. Rev. A **110**, 063724 (2024).

[50] J.-H. Wu, M. Artoni, and G. C. La Rocca, Non-hermitian degeneracies and unidirectional reflectionless atomic lattices, Phys. Rev. Lett. **113**, 123004 (2014).

[51] J.-H. Wu, M. Artoni, and G. La Rocca, Coherent perfect absorption in one-sided reflectionless media, Sci. Rep. **6**, 35356 (2016).

[52] J.-H. Wu, M. Artoni, and G. La Rocca, Perfect absorption and no reflection in disordered photonic crystals, Phys. Rev. A **95**, 053862 (2017).

[53] A. Mostafazadeh, Transfer matrix in scattering theory: A survey of basic properties and recent developments, Turk. J. Phys. **44**, 472 (2020).

[54] T. Inoue, N. Noguchi, M. Yoshida, H. Kim, T. Asano, and S. Noda, Unidirectional perfect reflection and radiation in double-lattice photonic crystals, Phys. Rev. Appl. **20**, L011001 (2023).

[55] H. Ramezani, Spectral singularities with directional sensitivity, Phys. Rev. A **103**, 043516 (2021).

[56] F. Zhang, N. S. Solodovchenko, H. Fan, M. F. Limonov, M. Song, Y. S. Kivshar, and A. A. Bogdanov, Non-hermitian singularities in scattering spectra of mie resonators, Sci. Adv. **11**, eadr9183 (2025).

[57] A. Mostafazadeh, Spectral singularities of complex scattering potentials and infinite reflection and transmission coefficients at real energies, Phys. Rev. Lett. **102**, 220402 (2009).

[58] A. Mostafazadeh, Optical spectral singularities as threshold resonances, Phys. Rev. A **83**, 045801 (2011).

[59] A. Mostafazadeh, Semiclassical analysis of spectral singularities and their applications in optics, Phys. Rev. A **84**, 023809 (2011).

[60] A. Mostafazadeh, Nonlinear spectral singularities of a complex barrier potential and the lasing threshold condition, Phys. Rev. A **87**, 063838 (2013).

[61] H. Ramezani, H.-K. Li, Y. Wang, and X. Zhang, Unidirectional spectral singularities, Phys. Rev. Lett. **113**, 263905 (2014).

[62] Z.-M. Wu, J.-H. Li, and Y. Wu, Vacuum-induced quantum-beat-enabled photon antibunching, Phys. Rev. A **108**, 023727 (2023).

[63] P.-O. Guimond, B. Vermersch, M. L. Juan, A. Sharafiev, G. Kirchmair, and P. Zoller, A unidirectional on-chip photonic interface for superconducting circuits, npj Quantum Inf. **6**, 32 (2020).

[64] B.-J. Li, Y.-L. Zuo, L.-M. Kuang, H. Jing, and C.-H. Lee, Loss-induced quantum nonreciprocity, npj Quantum Inf. **10**, 75 (2024).